\documentclass[aps,prb,reprint,superscriptaddress]{revtex4-1}
\usepackage{graphicx}
\usepackage[colorlinks=true,urlcolor=blue,linkcolor=blue,citecolor=blue]{hyperref}
\usepackage[utf8]{inputenc}

\usepackage{color}
\usepackage{amsmath}
\usepackage{braket}

\usepackage{natmove}

\begin{document}
\newcommand{\etal}{\emph{et~al.}}
\newcommand{\abinitio}{\emph{ab initio}}
\newcommand{\Abinitio}{\emph{Ab initio}}
\newcommand{\pawT}{\mathcal T}

\title{Localized surface plasmon resonance in silver
    nanoparticles: Atomistic first-principles time-dependent
    density-functional theory calculations}

\author{M. Kuisma}
\affiliation{Department of Physics, Tampere University of Technology, P.O. Box 692, FI-33101 Tampere, Finland}
\affiliation{Department of Microtechnology and Nanoscience, MC2, Chalmers University of Technology, SE-41296 G\"oteborg, Sweden}
\email{mikael.kuisma@chalmers.se}
\author{A. Sakko}
\author{T. P. Rossi}
\affiliation{COMP Centre of Excellence, Department of Applied Physics, Aalto University School of Science, FI-00076 AALTO, Finland.}
\author{A. H. Larsen}
\affiliation{Nano-Bio Spectroscopy Group and European Theoretical Spectroscopy Facility (ETSF), Universidad del Pa\'{i}s Vasco UPV/EHU,
Avenida de Tolosa 72, E-20018 Donostia-San Sebasti\'{a}n, Spain}
\author{J. Enkovaara}
\affiliation{CSC-IT Center for Science Ltd., P.O. Box 405, FI-02101 Espoo, Finland}
\affiliation{COMP Centre of Excellence, Department of Applied Physics, Aalto University School of Science, FI-00076 AALTO, Finland.}
\author{L. Lehtovaara}
\affiliation{Department of Chemistry, Nanoscience Center, University of Jyv\"askyl\"a, FI-40014 Jyv\"askyl\"a, Finland}
\author{T. T. Rantala}
\affiliation{Department of Physics, Tampere University of Technology, P.O. Box 692, FI-33101 Tampere, Finland}

\begin{abstract}
  We observe using \abinitio{} methods 
that localized surface plasmon resonances
  in icosahedral silver nanoparticles enter the asymptotic
  region already between diameters of 1--2~nm, converging close to
  the classical quasistatic limit around 3.4~eV. 
We base the observation on time-dependent density-functional theory simulations of
the icosahedral silver clusters Ag$_{55}$ (1.06~nm), Ag$_{147}$
    (1.60~nm), Ag$_{309}$ (2.14~nm), and Ag$_{561}$ (2.68~nm). 
The simulation method combines the adiabatic GLLB--SC exchange--correlation functional with real time propagation
  in an atomic orbital basis set using the projector
  augmented wave method. The method has been implemented to the electron structure code GPAW
  within the scope of this work. 
We obtain good agreement with experimental data and modelled results,
  including photoemission and plasmon resonance. 
Moreover we can extrapolate the \abinitio{} results to the classical quasistatically modelled icosahedral clusters.

\end{abstract}

\maketitle

\section{Introduction}

\noindent Localized surface plasmon resonances (LSPR) of silver nanoparticles (AgNPs) exhibit strong UV--VIS 
absorption. 
The LSPR can be tuned by fabrication techniques\cite{Jensen2000}, or by functionalization\cite{Tan2014}, 
and it is sensitive to the nanoparticle's environment\cite{Bingham2010}. Sensitivity and tunability of AgNPs can be utilized in
sensing\cite{Mayer2011}, surface-enhanced spectroscopies\cite{Willets2007}, plasmon-enhanced chemistry\cite{Xiao2013}, and photovoltaic 
applications\cite{Pillai2010}. Much of the wide interest in AgNPs originates from their role as building blocks of nanophotonic devices, 
such as optical nanoantennas\cite{Giannini2011}. The ability to predict the relation between their structure and 
operation is crucial for the applications. The optical characteristics of large noble metal NPs 
($>10$~nm) are well known, and their LSPR can be simulated using classical electromagnetic theory. 
For example, large spherical AgNPs have a LSPR at 355~nm (3.5~eV), whereas icosahedral particles are 
slightly redshifted and have broader absorption arising from several LSPR 
modes that overlap closely in energy.\cite{Kelly2003,Noguez2007} However, as the diameter of the NPs decreases, the LSPR 
blueshifts with the frequency being inversely proportional to the diameter\cite{Charle1998} 
and finally, the absorption spectrum changes to a typical cluster spectrum characterized by several 
individual transitions between quantized energy levels.\cite{Joswig2008,Xia2009} For diameters 
smaller than 10~nm, the sensitivity of the LSPR to the shape and surroundings of the AgNP becomes 
important which is reflected in the difficulty of interpretation of experiments.

Previous theoretical studies on the LSPR in AgNPs are limited to quantum mechanical calculations 
of small clusters\cite{LopezLozano2013} and jellium models\cite{Prodan2002}, or to classical 
electromagnetic theory for large NPs.\cite{Kelly2003} Between diameters of 1--5~nm, the classical 
electromagnetic theory does not provide an adequate description of the possible quantum effects
as it is scale invariant and therefore predicts no size dependence. Jellium models ignore --- or at best 
approximate\cite{Serra1997} --- the effect of d-electrons and atomic structure which are crucial 
to the proper description of AgNPs. \Abinitio{} methods are limited to small clusters:
Time-dependent\cite{PhysRevLett.52.997} (TD) density-functional theory\cite{PhysRev.136.B864,PhysRev.140.A1133} (DFT)
has been used to model Ag$_{55}$ and Ag$_{147}$\cite{LopezLozano2013}, as well as
nanoshells up to Ag$_{272}$.\cite{doi:10.1021/jp5016565}

Typically in such studies one uses the adiabatic local density approximation (ALDA) or adiabatic generalized gradient approximation (AGGA) as 
exchange--correlation (XC) functionals, even though LDA and GGA are known to predict a too high-lying 
d-electron band and therefore to severely overestimate the d-band screening.\cite{Yan2011} 
This results in decreased oscillator strength and lowered plasmonic frequency compared to experiments.
Both experimental\cite{Sonnichsen2002} and theoretical\cite{LopezLozano2014} works have confirmed that 
the position of the d-band strongly influences the plasmonic properties. Quantitative theory must 
be based on a more accurate description of the d-band. 

A recent experimental study of Scholl~\etal~found quantum effects influencing the optical 
properties of AgNPs with diameters as large as 10~nm.\cite{Scholl2012} In particular, their 
electron-energy-loss spectroscopic (EELS) measurements on NPs showed a 
significant (0.5~eV) blueshift of the LSPR when the diameter decreased from 7~nm to 2~nm. This 
disagrees with previous experimental results for freestanding 
clusters\cite{Tiggesbaumker1993,Charle1998}, and Haberland has suggested that the blueshift is 
not due to the quantum effects but due either to the interaction of the LSPR with the substrate or the residual 
ligand molecules.\cite{Haberland2013} This controversy exemplifies that without tools that can 
simulate the optical properties of NPs from molecular size up to the classical limit, it 
is difficult to separate the quantum effects from other factors.

In this work, we present an \abinitio{} theoretical analysis of freestanding AgNPs up to diameter of 2~nm, and show that it is unlikely 
that the results of Scholl \emph{et al.} correspond to freestanding AgNPs.
Using atomistic first-principles 
calculations in the TDDFT framework, we are able to 
obtain the macroscopic LSPR already at a diameter of 2~nm. Our calculations show that the resonance shifts only 
by~0.2~eV above that.
Our results agree with the experimental cluster data for both the smallest and the 
largest structures.
Concurrently with explaining the experimental findings, we show that accurate treatment of 
interband (d-electron) excitations is crucial for a reliable description of AgNP plasmonics. Therefore, we recommend
the adiabatic Gritsenko--van Leeuwen--van Lenthe--Baerends---solid-correlation potential (GLLB--SC) \cite{Kuisma2010} for approximating the exchange and correlation effects
for the optical properties of noble metal NPs. The potential is
a modification of the GLLB-potential \cite{Gritsenko1995} to be better suited
for solids and surfaces and with added correlation.

In Section~\ref{sec:methods} we describe the details of our implementation of linear combinations of atomic orbitals with time-dependent
density-functional 
theory (LCAO-TDDFT), and elaborate the relevance of the GLLB--SC potential for the proper description of plasmonics.
In Section~\ref{sec:model} we give basic background information about the quantum mechanical and the 
electrodynamical model.  In Section~\ref{sec:results} we analyse the obtained results, and compare them to experimental EELS and photoemission
data. In Section~\ref{sec:accuracy} we carefully benchmark the accuracy of our method.
In Section~\ref{sec:conclusions} we summarize the results and discuss the relevance of proper Kohn--Sham eigenvalue description
for accurate absorption spectra in AgNPs.


\section{Methods}\label{sec:methods}

\noindent
The main computational challenges in simulating the photoabsorption spectrum of nanoplasmonic structures using TDDFT are
1) quality of the XC functional, especially for the description of the silver d-band, 
2) the numerical discretization scheme for the wavefunctions and the density, which must be flexible enough to describe the LSPR, and
3) the method for optical properties must be fast, parallelizable, and scale well with respect to system size. 
Each of the challenges will be addressed in the following subsections.

\subsection{Time-dependent Density-Functional Theory}
\noindent
Time-dependent density-functional theory is a well established tool for calculating electronic excitations.
As in DFT, the most crucial aspect of TDDFT is the exchange--correlation potential, which is time-dependent in this case.
The time-dependent Kohn--Sham equations for the electronic orbitals $\Psi_i$ are
\begin{equation}
\left( -{\rm i} \partial_t - \frac{1}{2}\nabla^2 + v_{\rm KS}[n({\bf r}, t)]({\bf r}, t) \right) \Psi_{i}({\bf r},t) = 0,
\label{eq:tdks}
\end{equation}
\noindent
where $v_{\rm KS}$ is the Kohn--Sham potential, the time-dependent density is given by
\begin{equation}
n({\bf r}, t) = \sum_i f_i |\Psi_i({\bf r}, t)|^2,
\end{equation}
and $f_i$ are the occupation numbers of the orbitals.
 
In the general formalism, the exchange--correlation part of the Kohn--Sham potential $v_{\rm xc}$ depends causally on all previous 
densities. In a practical and widely used adiabatic approximation, the potential depends only on the instantaneous density.
We will use this approximation also in the case of the GLLB--SC potential, with one further modification, as discussed
in the next section.

\subsection{Adiabatic GLLB--SC}
\noindent
Adiabatic (semi)local density approximations, such as ALDA and AGGA, are applicable for nearly free-electron metals, but for noble metals 
the situation is different because they overestimate the polarizability of d-electrons.
This is due to their Kohn--Sham spectrum, since they predict too delocalized d-band in addition
it being too shallow \cite{Yan2011,PhysRevB.86.241404}.
To overcome this problem we employ the adiabatic GLLB--SC
potential\cite{Gritsenko1995,Kuisma2010} that includes the exchange-hole and correlation potential of the Perdew--Burke--Ernzerhof functional
for solids and surfaces (PBEsol) \cite{Perdew2008}, and is additionally supplemented by a computationally efficient approximation 
of the hole response part (see, e.g., Ref.~\onlinecite{QUA:QUA5}) of the exact-exchange optimized effective potential.\cite{Staedele1997}

GLLB--SC introduces an orbital energy dependent localization of the exchange hole which reduces 
self-interaction and yields better asymptotic behavior than LDA or GGA\cite{Kuisma2010}. So far GLLB--SC has been mostly applied for 
predicting semiconductor band gaps\cite{Kuisma2010,Castelli2012}, but recently Yan~\etal~showed that it also 
yields good results for Ag surface plasmons because of the improved d-band description.\cite{Yan2011} We 
employ this finding, but extend it further by applying GLLB--SC also for the dynamic response (in 
Ref.~\onlinecite{Yan2011}, GLLB--SC was used only for the ground state whereas the linear response calculation 
employed ALDA). 

We obtain an adiabatic GLLB--SC approximation by replacing the time-dependent response coefficients
$w_i(t)$ with their
time-independent ground-state values $w_i(t=0) = K_g \sqrt{\epsilon_f - \epsilon_i}$ in the GLLB--SC potential
(see Eqns.~(16) and (22) of Ref.~\onlinecite{Kuisma2010}):
\begin{equation}
v_{\textrm{GLLB--SC}}({\bf r}) = v_{\textrm{x-hole}}({\bf r}) + \sum_i w_i(t=0) \frac{|\psi_i({\bf r})|^2}{n({\bf r})} + v_c({\bf r}),
\end{equation}
where $v_{\textrm{x-hole}}({\bf r})$ is the Coulomb potential due to
the exchange hole obtained from the exchange hole of PBEsol
evaluated at the instantaneous density,
$v_c({\bf r})$ is the semi-local PBEsol correlation potential, and the
remaining term is an approximation to the response of the Coulomb
potential of the exchange--correlation hole to density perturbations.
We choose $w_i(t)=w_i(t=0)$, since it is the simplest obtainable approximation
and computationally attractive.
It is plausible that this approximation is accurate in our simulations because
we apply a small perturbation which will not significantly change the density, and thus
not induce large oscillations of $w_i(t)$.
In addition, our preliminary adiabatic time-dependent
Krieger--Li--Iafrate (TD-KLI) \cite{PhysRevA.45.101} calculations
indicate that the effect of $w_i(t)$ compared to $w_i(t=0)$ in the systems considered here are negligible 
in the linear response regime.

It is difficult to estimate the effect of this approximation, or the effect of non-adiabatic exchange--correlation effects exactly.
However, there is much evidence from historical work that already the random phase approximation (pure Coulomb kernel) without any XC kernel
is sufficient to describe plasmonics.\cite{PhysRev.85.338}
Therefore, the adiabatic GLLB--SC approximation should be a sufficient description for AgNP plasmonics.

\subsection{Real Time Propagation with Basis Sets}
\noindent
The wavefunctions are represented as linear combinations of atomic orbitals (LCAO)
together with the projector augmented wave method\cite{PhysRevB.50.17953} (PAW) as implemented in
the GPAW package \cite{Enkovaara2010, Larsen2009}. The smooth pseudo wavefunctions are written 
as a linear combination
\begin{equation}
\tilde{\Psi}_i({\bf r},t) = \sum_\mu C_{\mu i}(t) \tilde{\phi}_\mu({\bf r}- {\bf R}^\mu)
\label{basis}
\end{equation}
\noindent
of atom centered orbitals $\tilde{\phi}_\mu({\bf r}- {\bf R}^\mu)$
with expansion coefficients $C_{\mu i}(t)$.
The PAW projection operator\cite{PhysRevB.50.17953} $\widehat \pawT$ can be used to reconstruct the all-electron functions 
as $\Psi_i({\bf r},t) = \widehat \pawT \tilde{\Psi}_i({\bf r},t)$.
The PAW form of the time-dependent Kohn--Sham equations \eqref{eq:tdks} is
\begin{equation}
\left[ \widehat \pawT^\dagger \left(-{\rm i}{\frac{\partial}{\partial t}}\right) \widehat \pawT + \widehat \pawT^\dagger \widehat H_{\rm KS}(t) \widehat \pawT \right] \tilde \Psi_i ({\bf r}, t) = 0,
\label{kspaw}
\end{equation}
\noindent
where $\widehat H_{\rm KS}(t)$ is the Kohn--Sham Hamiltonian for non-interacting electrons.

Substituting Eq.~\eqref{basis} into Eq.~\eqref{kspaw} and multiplying with $\int {\rm d}{\bf r} \tilde{\phi}_\mu({\bf r})$ from the left, 
the equation can be cast into a matrix form
\begin{equation}
{\rm i} {\bf S} \frac{\mathrm d {\bf C}(t)}{\mathrm d t} = {\bf H}(t) {\bf C}(t),
\end{equation}
\noindent
with the overlap matrix $S_{\mu\nu} = \braket{\tilde \phi_\mu|\widehat \pawT^\dagger \widehat \pawT |\tilde \phi_\nu}$
and the Hamiltonian matrix $H_{\mu\nu}(t) = \braket{\tilde \phi_\mu|\widehat \pawT^\dagger \widehat H_{\rm KS}(t) \widehat \pawT|\tilde \phi_\nu}$.
$\mathbf C(t)$ is the matrix of LCAO expansion coefficients $\{C_{\mu i}(t)\}$ defined in Eq.~\eqref{basis}.
The overlaps $S_{\mu\nu}$ and the projection operator $\widehat \pawT$ are constant because the nuclei are assumed to be
stationary.

In this approach, the time-dependent density and potential are expressed on a uniform grid,
and the matrix elements of the potential are evaluated on this grid.\cite{Larsen2009} 
The smoothness of these quantities allows for a very coarse grid spacing, and the
LCAO-PAW pseudo wavefunctions form a small, local, and efficient representation suitable for systems with hundreds of atoms.\cite{Larsen2011}

We calculate the optical absorption spectrum of AgNPs using 
the time-propagation (TP) approach to TDDFT\cite{Yabana1996,:/content/aip/journal/jcp/128/24/10.1063/1.2943138}. The greatest advantage of TP-TDDFT is the scaling of the computational 
requirements with respect to system size compared to other methods, such as Casida's approach\cite{Casida1995}. Despite its better 
scaling, the large prefactor has so far limited the applicability of the TP-TDDFT approach. 

Following the TP-TDDFT procedure for the optical response\cite{Yabana1999}, we here excite the system by an
instantaneous electric field  ${\bf E}({\bf r}, t)=E_0
{\bf\hat{e}}_{\rm tt} \delta(t)$, where the field strength
$E_0=0.0001$~a.u.\ is sufficiently small to avoid nonlinear effects,
and the direction ${\bf\hat{e}}_{\rm tt}$ of the electric field 
is chosen to be from tip to tip, i.e., along the five-fold symmetry
axis of the icosahedron.
The optical absorption spectrum is obtained by Fourier transforming the induced dipole moment along the excitation axis.\cite{Yabana1996}

After the initial kick, the propagation is performed with a reliable and numerically stable semi-implicit Crank--Nicolson method.
In brief, the method can be described as follows. In the prediction step, we solve
\begin{align}
\left({\bf S} + {\rm i} \frac{{\rm d} t}{2} {\bf H}(t)\right) {\bf C}'(t+{\rm d}t)
= \left({\bf S} - {\rm i} \frac{{\rm d} t}{2} {\bf H}(t)\right) {\bf C}(t),
\end{align}
\noindent
for $\mathbf C'(t + {\rm d} t)$,
where ${\bf S}$ and ${\bf H}(t)$ are the basis set representations of the PAW overlap and Hamiltonian operator respectively.
The operations are parallel with matrices being distributed using ScaLAPACK\cite{scalapack} and BLACS\cite{blacs}.
After obtaining the initial approximation for the wavefunctions, 
the predict--correct method is applied. We then obtain an estimate for the
Kohn--Sham Hamiltonian (including the XC potential) at the middle of the time step%
\begin{equation}
{\bf H}(t+{\rm d}t/2) \approx ({\bf H}(t) + {\bf H}'(t+{\rm d}t))/2,
\label{eqa}
\end{equation}
where $\mathbf H'(t + \mathrm d t)$ is evaluated from $\mathbf C'(t + \mathrm dt)$,
and then propagate the wavefunction to $t+\mathrm d t$ in the correction step which solves
\begin{align}
\left({\bf S} + {\rm i} \frac{{\rm d} t}{2} {\bf H}(t+{\rm d}t/2)\right) {\bf C}(t+\mathrm d t)\nonumber\\
=\left({\bf S} - {\rm i} \frac{{\rm d}t}{2} {\bf H}(t+{\rm d}t/2)\right) {\bf C}(t),
\end{align}
for $\mathbf C(t+\mathrm d t)$.
This results in $\mathcal O(N^3)$ scaling with respect to the number of electrons in the system,
to be compared to the GPAW's Casida implementation of $\mathcal O(N^5)$ or 
the real-space time propagations $\mathcal O(N^2)$. However, the constant factor in the grid propagation is so large,
that our scheme performs 1 to 3 orders of magnitude faster on systems of several thousand electrons.
The timings for propagation are indicated in Table~\ref{perf}.
\begin{table}[htp]
\centering
\caption{Performance of the LCAO-TDDFT code for
time propagation of AgNPs
for 1000 time steps of duration 10 as.}
\begin{tabular}{|l|rrrrr|}
\hline
System & Cores & Wall hrs & CPU hrs & Electrons & Basis functions \\
\hline
\hline
Ag$_{55}$ & 64 & 4.5 & 288 & 605 & 990 \\
Ag$_{147}$ & 64 & 18.0 & 1152 & 1617 & 2646 \\
Ag$_{309}$ & 256 & 28.5 & 7296 & 3399 & 5562 \\
Ag$_{561}$ & 512 & 42.0 & 21504 & 6171 & 10098 \\
\hline
\end{tabular}
\label{perf}
\end{table}

\section{Model}\label{sec:model}
\noindent Both classical and quantum mechanical models are employed in this work. The quantum mechanical model is 
atomistic and \abinitio, relying only on DFT and TDDFT calculations. The plasmonic peak in our 
quantum mechanical model depends both on shape and size effects.  The classical model
is based on empirical dielectric functions and can only model shape effects. However,
in the large particle limit, the two methods should agree.
Therefore, we can test the performance of our computational model also by extrapolating to macroscopic Mie scattering limit.

In both models we consider icosahedral clusters.  Charged Ag$_{55}$ has been experimentally identified as 
icosahedral\cite{Schooss2005}, and we choose to keep the icosahedral geometry to avoid shape effects even 
though the minimum energy structure is expected to change for larger clusters\cite{Baletto2002}. The difference between 
icosahedron and sphere in the macroscopic limit is well understood\cite{Kelly2003,Noguez2007} and does not influence our conclusions. The 
atomic structure of icosahedral Ag$_{55}$ used in our calculations includes a central atom with two 
icosahedral Mackay layers. We create clusters up to Ag$_{561}$ by adding Mackay layers one by one, using a 
bond length of 3.0~Å. The ideal icosahedral clusters are then relaxed with the 
LDA functional. In these calculations the grid spacing is 0.2~Å, and the size of the cubic cell 
is chosen so that all atoms are at least 5.0~Å away from the cell boundary.
We use the default double-$\zeta$ polarized (DZP) basis set provided with GPAW for the geometry relaxations\cite{Larsen2009}. 

TDDFT simulations are performed for 30~fs using time steps of 10~as.  These calculations use a coarse grid spacing of 0.3~Å
and an expanded atomic basis set; we will further discuss these parameters in Section \ref{sec:accuracy}.
All spectra are calculated using a Gaussian broadening of 0.16~eV FWHM.

In the following we consider classical electrodynamic approximations.
First, the photoabsorption of a spherical NP of volume $V$ is given by the quasistatic limit of Mie theory:
\begin{equation}
    S(\omega) = \frac{3V\omega}{2\pi^2}\mbox{Im}\left[\frac{\epsilon(\omega)-1}{\epsilon(\omega)+2}\right].
    \label{eq:Mie}
\end{equation}
By using the experimentally determined permittivity $\epsilon(\omega)$ for silver presented in Ref.~\onlinecite{Johnson1972}, Eq.~\eqref{eq:Mie} yields a strong LSPR at 3.5~eV.
For more complicated shapes, such as icosahedra, one has to employ computational electrodynamics.
In this work we use a quasistatic (QS) version\cite{Coomar2011} of the widely used finite-difference-time-domain (FDTD) method, as implemented in GPAW\cite{Sakko2014}.
Like in photoabsorption calculations with TDDFT, in the QSFDTD method one perturbs the system by an external field and analyses the time-dependent dipole moment.
The frequency-dependent dielectric permittivity of classical material is approximated using a set of Lorentzians.
To obtain an accurate representation of the dielectric function of Ag especially near the LSPR, we start from the parametrization presented in Ref.~\onlinecite{Coomar2011} which uses 9 Lorentzians, add one extra Lorentzian, and refit the dielectric function against the experimental data\cite{Johnson1972} with weight function $w(\omega)={\rm exp}(-(\omega/\mathrm{eV}-3.5)^2)$.

The QSFDTD calculations are performed using a regular grid of $96\times 96\times 96$ points.
Since this method is 
size invariant, only a particle shape needs to be specified.
We thus specify the shape as an icosahedron with a length of 40 points along its axis, securing adequate surrounding vacuum.
The material is represented by a mask which assigns a value of either 1 (material) or 0 (vacuum) to each point.
To ensure high numerical accuracy of the finite-difference operators,
we smoothen the edge of the icosahedron artificially over 2--3 grid points along the faces so that points along
the faces are effectively a mixture of vacuum and silver.



\section{Results}\label{sec:results}

\begin{figure}[htp!]
  \includegraphics[width=3.3in]{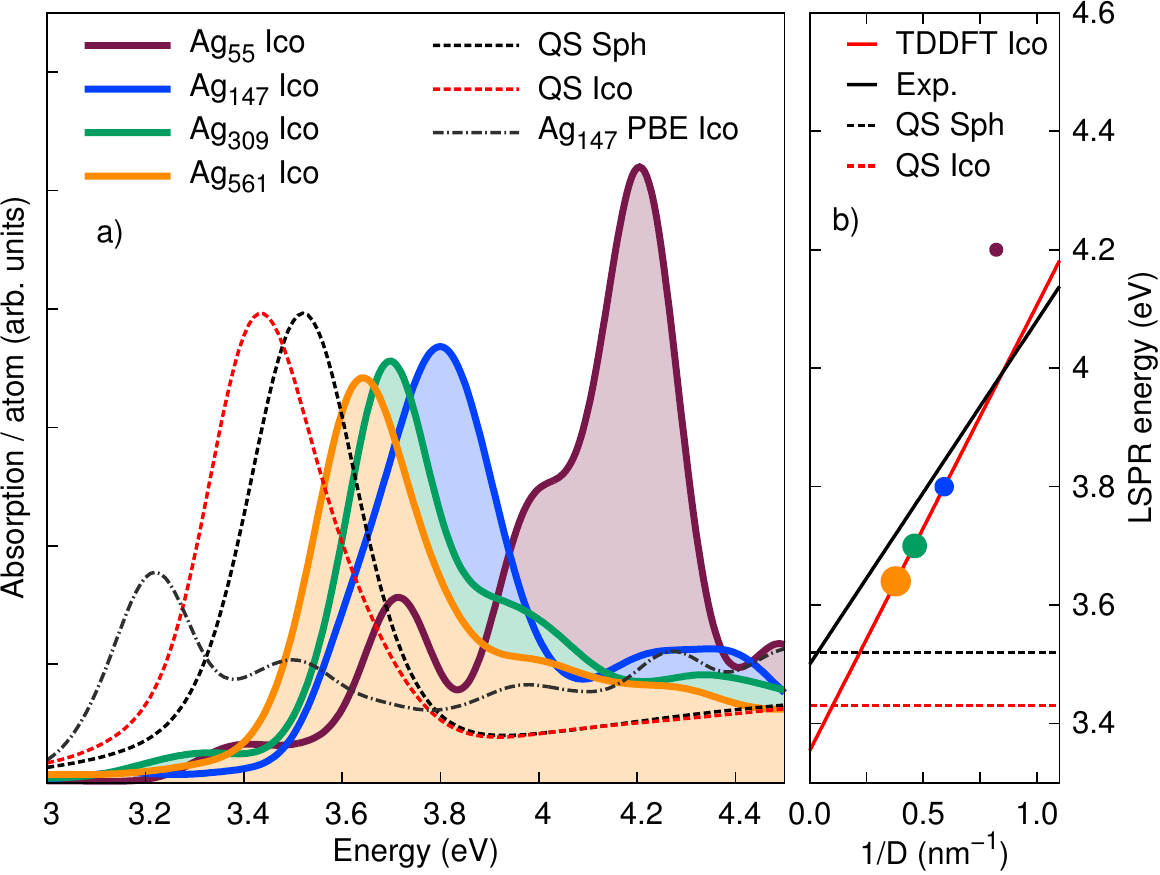}
  \caption{a)
Photoabsorption spectra for AgNPs normalized by number of atoms.
    The spectra for icosahedral Ag$_{55}$, Ag$_{147}$, Ag$_{309}$ and Ag$_{561}$ are calculated with adiabatic GLLB--SC TDDFT.
    For comparison, the PBE-calculated spectrum of Ag$_{147}$ is also shown.
    Classical QSFDTD spectra are shown for spherical (QS Sph) and icosahedral (QS Ico) AgNPs,
    calculated with the empirical dielectric function. These spectra are normalized using the empirical silver density and thus
    the peak strengths are directly comparable.
b)  LSPR energy (circles) of icosahedral AgNPs as a function of inverse particle diameter with GLLB--SC TDDFT.
    Solid black and red lines correspond to the experimental data for spherical NPs\cite{Charle1998} and a linear fit to our results, respectively.
    Dashed horizontal lines represent classical limits (QSFDTD) for spherical (QS Sph) (3.52~eV) and icosahedral (QS Ico) (3.43~eV) NPs.}
\label{fig1}
\end{figure}

\noindent Fig.~\ref{fig1}(a) shows the GLLB--SC TDDFT absorption spectra of icosahedral Ag$_{55}$, Ag$_{147}$, Ag$_{309}$, and Ag$_{561}$
clusters divided by number of atoms in the system. For comparison, we present classical QSFDTD results for icosahedral (dashed red) and spherical shape (dashed black).
These correspond to the limit of large clusters as given by the quasistatic approximation. 
Fig.~\ref{fig1}(b) shows excitation energies of absorption peaks with respect to the inverse diameter of the cluster.
For NPs larger than Ag$_{55}$, the excitation energies of the most intense
peak as a function of inverse diameter lie on a line (solid red) which
extrapolates to 3.35~eV in the large particle limit,
very close to the mesoscopic limit for
{\em icosahedral} AgNPs at 3.43~eV (dashed red) obtained from the QSFDTD calculation.
The agreement of the quasistatic mesoscopic limit with the quantum mechanical asymptotic limit suggests that the quantum mechanical model correctly describes the shape effect.
For comparison, a linear fit to experimental data (solid black) is shown for
{\em spherical} AgNPs in argon matrix\cite{Charle1998} and also the mesoscopic limit for spherical NPs from the QSFDTD calculation (dashed black).
The experimental values are shifted to the vacuum LSPR value of 3.5~eV to account for the Ar matrix as suggested by Haberland\cite{Haberland2013}.
The experimental and the simulated data show remarkable agreement, both in the asymptotic limit
and in the size dispersion. The differences can be attributed to slightly different AgNP shape and structure.
These observations suggest that the quantum mechanical model describes the finite size effect in the AgNP plasmonics well.

In Fig.~\ref{fig1}(a), in addition to the LSPR energy, also the area of the plasmon peak per particle (oscillator strength) agrees 
well with the classical electrodynamics simulation (dashed red).
These observations strongly indicate that (i) adiabatic GLLB--SC provides realistic d-band screening
in Ag nanostructures, and (ii) the macroscopic size range is reached for AgNPs of diameter $\sim$2~nm. 
In Fig.~\ref{fig1}, for comparison, we have included the spectrum of Ag$_{147}$ calculated with PBE.\cite{PhysRevLett.77.3865}
Importantly, the previous conclusions cannot be drawn from APBE calculations 
because they underestimate the LSPR energy by $\sim$0.5~eV, and greatly underestimate the intensity as seen on
Fig.~\ref{fig1}(a).

\begin{figure}[ht!]
\includegraphics[width=3in]{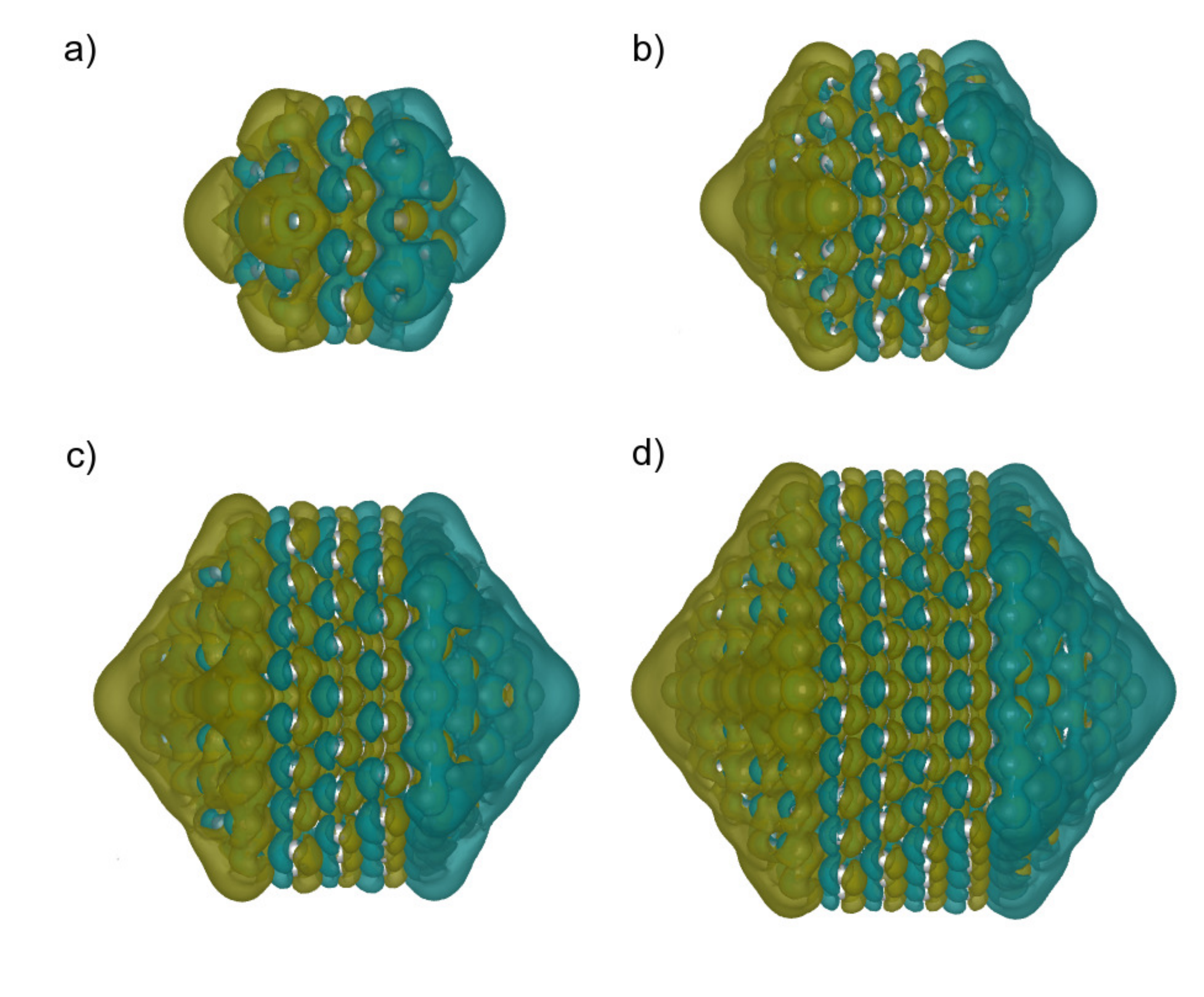}
\caption{Calculated induced electron densities of LSPR for a) Ag$_{55}$, b) Ag$_{147}$, c) Ag$_{309}$, and d) Ag$_{561}$.
The Ag sp-band forms a localized surface plasmon on the surface of the cluster, whereas the d-electrons polarize in the opposite direction.
}
\label{induced}
\end{figure}

Previous works\cite{Bernadotte2013, Malola2013} have demonstrated the importance of visual interpretation for characterizing the LSPR in molecules and NPs.
The induced electron densities of LSPRs in Ag$_{55}$, Ag$_{147}$, Ag$_{309}$, and Ag$_{561}$
are shown in Fig.~\ref{induced}.
The exact quantity shown is the transition density
at the plasmon frequency $\omega$ of each AgNP, i.e., a sine transform
\begin{align}
  \tilde n(\mathbf r, \omega) = \int_0^{\infty} \mathrm dt \, [n(\mathbf r, t) - n(\mathbf r, 0)]\ \mathrm e^{-\sigma^2 t^2 / 2} \sin \omega t
\end{align}
of the charge density fluctuation.
The damping is 
given by $\sigma_{\rm FWHM} = 2 \sqrt{2 \log 2}\, \sigma = 0.16~\mathrm{eV}$.
We observe that
the Ag sp-band near the Fermi energy forms a localized surface plasmon mainly at the two opposing sides of the icosahedron, 
whereas d-electrons polarize in the opposite direction and thus create a counteracting screening field at the central region.
This screening is overestimated by PBE, causing the drop in plasmon energy and intensity.
The figure corresponds to the classical picture of plasmons as a charge cloud oscillating between the opposite sides of the AgNP.
The visual inspection thus supports our finding that the macroscopic plasmon forms in the clusters of this size range. 

\begin{figure}[htb]
\includegraphics[width=3.3in]{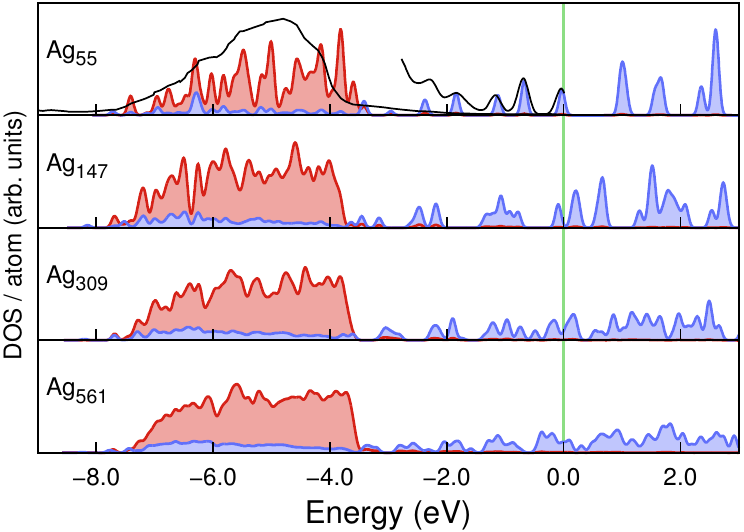}
\caption{Density of states of icosahedral AgNPs calculated with GLLB--SC, projected separately for d-band (red) and sp-band (blue).
The black curves are experimental data\cite{Hakkinen,Wortmann}, shifted to align the Fermi levels. The charge state of Ag$_{55}$ has insignificant
effect on the quantitative agreement.}
\label{dos}
\end{figure}

Fig.~\ref{dos} shows the experimental photoemission data from two sources\cite{Hakkinen,Wortmann} on Ag$_{55}$ compared to sp and d-band 
projected local density of states of the quantum mechanical clusters. The d-band position of GLLB--SC matches well with the 
experimental data, as has also been observed earlier\cite{Yan2011}.  In addition, the superatom shell description is in quantitative agreement
with photoemission data.\cite{Hakkinen}

\section{Accuracy of the method}\label{sec:accuracy}
\noindent
The PAW dataset used to represent Ag includes the 5s and 4d orbitals as valence states, and is based
on the default parameters of GPAW for the 11-electron Ag setup (e.g., the PBE Ag setup from GPAW-setups v0.8.7929)
but generated with the GLLB--SC functional.

In GPAW, one commonly uses a DZP numerical basis set to represent
the wavefunctions.\cite{Larsen2009}  This basis set includes the atomic Kohn--Sham orbital for each occupied valence state,
one extra radial function for each atomic KS orbital generated using the standard ``split-valence'' scheme in GPAW,
plus a polarization function which for transition metals is p-type.
For the details on the construction of the basis sets, see Ref.~\onlinecite{Larsen2009}. 
This basis set is designed for ground-state calculations and would not be expected to
(and indeed does not)
accurately predict properties that depend of unoccupied states.
To better represent the effect of the unoccupied 5p orbitals, we replace the standard p-type polarization function
with the actual Kohn--Sham orbital of the 5p state plus its usual split-valence function.

\begin{figure}[h!]
\centering
\includegraphics[width=\columnwidth]{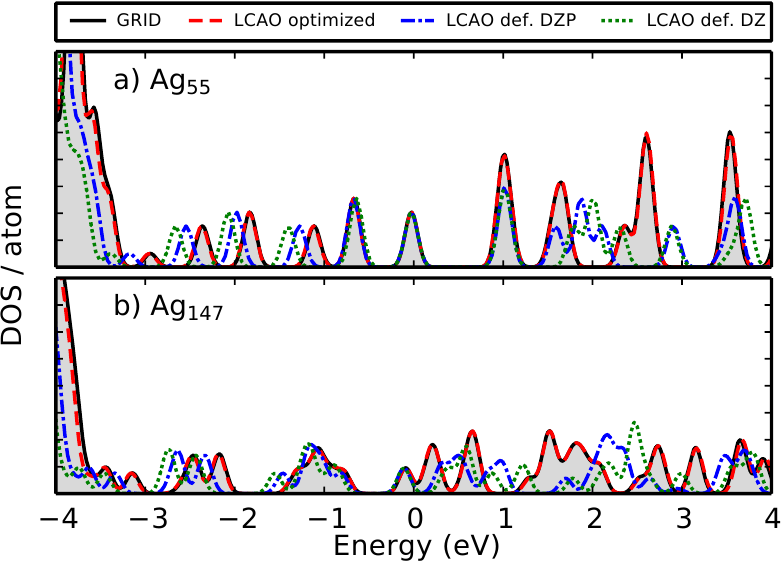}
\caption{The density of states of (a) the Ag$_{55}$ cluster and (b) the Ag$_{147}$ cluster calculated with different LCAO basis sets as well as grid mode.}
\label{fig dos}
\end{figure}

Outside of this, we use the specific generation parameters\cite{Larsen2009} of 0.07 eV
confinement energy to localize the KS orbitals and a tail norm of 0.2
to define the range of the split-valence functions.
These latter parameters we have optimized to provide an accurate density of states (DOS) in Ag$_{55}$
as compared to an accurate real-space grid calculation, but this optimization has very little effect compared to the
inclusion of the diffuse 5p valence orbital.  A comparison of DOSs is presented in Fig.~\ref{fig dos}.
We observe that without the diffuse 5p valence orbital the basis set is not able to reproduce the correct DOS accurately,
particularly for high energies.

\begin{figure}[h!]
\centering
\includegraphics[width=.95\columnwidth]{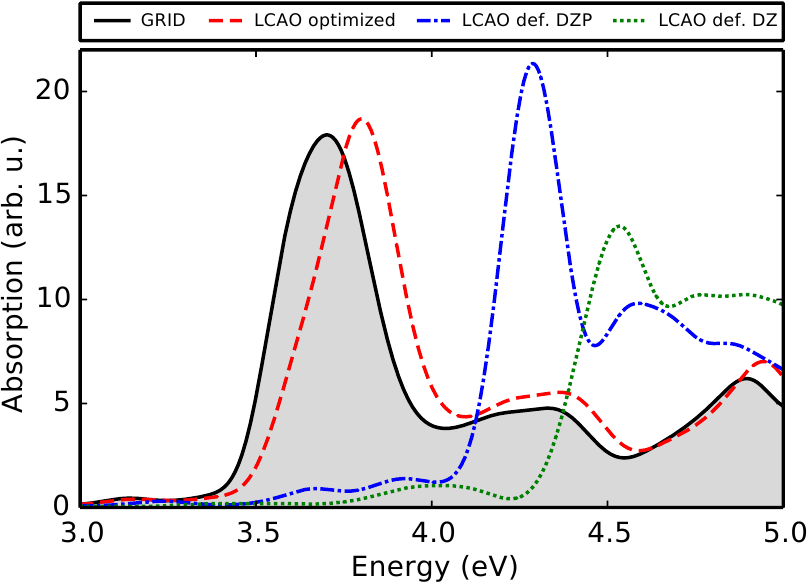}
\caption{The photoabsorption spectrum of the Ag$_{147}$ cluster calculated with different LCAO basis sets 
and the grid mode.}
\label{fig:lcao_spec}
\end{figure}

Fig.~\ref{fig:lcao_spec} presents the photoabsorption spectrum of the Ag$_{147}$ cluster calculated with different basis sets and on a real-space grid. Like in the DOS comparison, we note that the enhanced basis yields significantly better agreement with the grid mode than the default basis sets. In comparison to the real-space calculation, the enhanced LCAO basis reproduces the spectrum within $\sim{}\!$0.1~eV and $\sim{}\!$5\% accuracy for peak energy and intensity, respectively.  This approach yields a transferable basis set that can be expected to describe both DOS and the optical response of larger 
clusters with good accuracy.  To obtain further improvement in accuracy, more elaborate approaches can be used to enhance the basis set.\cite{Rossi2015}

To obtain 
good convergence with respect to the vacuum size, it is essential 
not to use zero boundary conditions for solving the Hartree potential of the Poisson equation. In the current work, we employ a
multipole moment expansion \cite{Castro2003} in order to obtain the correct boundary values of the Hartree potential.  This allows us to 
scale down the required amount of vacuum from 15~Å to 5~Å and obtain significant speedup.

As indicated by Table~\ref{perf}, our method achieves good parallel scaling in
the weak sense, i.e., the computational time can be kept within
reasonable limits by increasing the number of CPU cores as the system
size increases. 

\section{Conclusions}\label{sec:conclusions}
\noindent
We have established and carefully benchmarked a real time propagation method using atomic basis sets to obtain accurate plasmonics as demonstrated here
for icosahedral silver clusters. The implementation is part of the free open-source GPAW package.
We have shown that the eigenvalue spectrum of the GLLB--SC potential matches the available experimental
photoemission data for icosahedral silver clusters
and that the method provides an accurate description of the plasmonic response in TDDFT calculations.

The observation that only the LSPR of Ag$_{55}$ does not fit the asymptotic line in Fig.~\ref{fig1}(b) suggests that the macroscopic regime is reached already at Ag$_{147}$.
However, comparison of the spectrum of Ag$_{147}$ with the larger clusters shows that the shape of the LSPR peak deviates from the larger clusters.
These quantum effects disappear for Ag$_{309}$ and larger clusters.
This threshold size for asymptotic LSPR behavior is remarkably small and agrees with experimental observations\cite{Charle1998,Haberland2013} as well as with simulations of monolayer protected Au clusters\cite{Malola2013}.

The impact of this study is threefold. Firstly, we show using \abinitio{}
simulations that the LSPR frequencies and intensities in icosahedral
AgNPs enter an asymptotic region already around the
diameter of 2~nm. The optical response converges close to
the classical limit of 3.43~eV for icosahedral AgNPs.
Our simulations are in good agreement with the experimental data, and the conclusion is further supported by visual examination and analysis of the DOS.
The presented results thus set the benchmark for the plasmonics of AgNPs, and explain the controversy between the recent EELS results with previous cluster experiments. 
\cite{Scholl2012, Haberland2013}
Secondly, the results show that adiabatic GLLB--SC provides an accurate
description of d-band screening in Ag nanostructures with
computational effort that is comparable to ALDA and AGGAs.
The final point of the study --- with probably the greatest impact in the long run --- is the efficiency of the combination of TP-TDDFT, LCAO and the PAW method.
The method is not limited to pure Ag nanostructures. Our
preliminary results show that it is also applicable to intermetallic nanostructures, such as Au--Ag core--shell
NPs, as well as to nanostructures with molecular parts, e.g., ligand protected AuNPs\cite{Malola2013} and metallic nanoantennas connected by molecular tunnel junctions\cite{Tan2014}. 

Altogether the combination of adiabatic GLLB--SC, LCAO-PAW with
extended basis, and the time-propagation method allows for accurate
simulations of LSPRs in noble metal nanostructures towards macroscopic sizes.

\section{Acknowledgments}
\noindent
We thank the Academy of Finland for financial support through Projects No.~269402 and No.~273499, through its Centres of Excellence Programme (2012--2017) under Project No.~251748, and through its National 
Graduate School of Materials Physics. T.P.R.\ acknowledges financial support from the Vilho, Yrj\"o and Kalle V\"ais\"al\"a Foundation. 
We thank CSC - IT Center
for Science Ltd.\ (Espoo, Finland) and the Aalto Science-IT project for computational resources. A.H.L.\ acknowledges support from the European Research Council 
Advanced Grant DYNamo (ERC-2010-AdG Proposal No.\ 267374) and Grupos Consolidados UPV/EHU del Gobierno Vasco (Grant No.~\mbox{IT-578-13}).


\bibliographystyle{apsrev4-1} 
\bibliography{sources}
\end{document}